\documentclass[preprint]{aastex}

\newcommand{\kepler}{{\it Kepler\/}}
\newcommand{\Kepler}{{\it Kepler\/}}
\newcommand{\prsa}{Pr\v{s}a}

\newcommand{\D}{{\tt D}}
\newcommand{\SD}{{\tt SD}}
\newcommand{\OC}{{\tt OC}}
\newcommand{\ELV}{{\tt ELV}}
\newcommand{\UNC}{{\tt UNC}}

\newcommand{\CAT}{{\tt CAT}}
\newcommand{\KOI}{{\tt KOI}}
\newcommand{\NEW}{{\tt NEW}}

\shorttitle{Kepler EB Catalog: Q2 Update}
\shortauthors{Slawson, \prsa, et al.}

\begin{document}

\title{{\em Kepler} Eclipsing Binary Stars. II.  2165 Eclipsing Binaries in the Second Data Release}

\author{Robert W.\ Slawson}
\affil{SETI Insitute, 189 Bernardo Ave., Mountain View, CA 94043}
\email{rslawson@seti.org}

\author{Andrej \prsa}
\affil{Villanova University, Dept. of Astronomy and Astrophysics, 800 E Lancaster Ave, Villanova, PA 19085}
\email{andrej.prsa@villanova.edu}

\author{William F.\ Welsh and Jerome A.\ Orosz}
\affil{San Diego State University, 5500 Campanile Dr., San Diego, CA 92182}

\author{Michael Rucker and Natalie Batalha}
\affil{San Jose State University, One Washington Square, San Jose, CA 95192}

\author{Laurance R.\ Doyle}
\affil{SETI Insitute, 189 Bernardo Ave., Mountain View, CA 94043}

\author{Scott G.\ Engle, Kyle Conroy, and Jared Coughlin}
\affil{Villanova University, Dept. of Astronomy and Astrophysics, 800 E Lancaster Ave, Villanova, PA 19085}

\author{Trevor Ames Gregg, Tara Fetherolf, Donald R.\ Short, and Gur Windmiller}
\affil{San Diego State University, 5500 Campanile Dr., San Diego, CA 92182}

\author{Daniel C.\ Fabrycky}
\affil{UCO/Lick, University of California, Santa Cruz, CA 95064}

\author{Steve B.\ Howell}
\affil{National Optical Astronomical Observatory, Tucson, AZ 85726}

\author{Jon M.\ Jenkins}
\affil{SETI Institute/NASA Ames Research Center, Moffett Field, CA 94035}
% 650-604-1111, jon.jenkins@nasa.gov 

\author{Kamal Uddin}
\affil{Orbital Sciences Corporation/NASA Ames Research Center, Moffett Field, CA 94035}
% 650-604-6782, akm.uddin@nasa.gov

\author{F.\ Mullally, Shawn E.\ Seader, and Susan E.\ Thompson}
\affil{SETI Institute/NASA Ames Research Center, Moffett Field, CA 94035}
% 650.604.3517, fergal.mullally@nasa.gov, SETI Institute/NASA Ames Research Center, Moffett Field, CA 94035     
% Shawn E. Seader, 650.604.4241, shawn.seader@nasa.gov SETI Institute/NASA Ames Research Center, Moffett Field, CA 94035
% Susan E. Thompson, 650.604.4249, susan.e.thompson@nasa.gov, SETI Institute/NASA Ames Research Center, Moffett Field, CA 94035

\author{Dwight T.\ Sanderfer}
\affil{NASA Ames Research Center, Moffett Field, CA 94035}  
% 650-604-3452, dwight.t.sanderfer@nasa.gov

\and
\author{William Borucki and David Koch}
\affil{NASA Ames Research Center, Moffett Field, CA 94035}

\begin{abstract}

The \Kepler\ Mission provides nearly continuous monitoring of
$\sim156\,000$ objects with unprecedented photometric precision.
Coincident with the first data release, we presented a catalog of 1879
eclipsing binary systems identified within the 115 square degree
\Kepler\ FOV.  Here, we provide an updated catalog augmented with the
second \Kepler\ data release which increases the baseline nearly
4-fold to 125 days.  386 new systems have been added, ephemerides and
principle parameters have been recomputed.  We have removed 42
previously cataloged systems that are now clearly recognized as short-period
pulsating variables and another 58 blended systems where we have
determined that the \Kepler\ target object is not itself the eclipsing
binary.
A number of interesting objects are identified.
We present several exemplary cases:
4 EBs that exhibit extra (tertiary) eclipse events;
and 8 systems that show clear eclipse timing variations
indicative of the
presence of additional bodies bound in the system.
We have updated
the period and galactic latitude distribution diagrams. With these
changes, the total number of identified eclipsing binary systems in
the \Kepler\ field-of-view has increased to 2165, 1.4\% of the
\Kepler\ target stars.

An online version of this catalog is maintained at
\anchor{http://keplerEBs.villanova.edu}{http://keplerEBs.villanova.edu}.
\end{abstract}


\section{Introduction}

The NASA \Kepler\ Mission, launched in March 2009, continues to
photometrically monitor $\sim156\,000$ targets within an 115 square degree
field in the direction of the constellation Cygnus.  Details and
characteristics of the \Kepler\ photometer and observing program have been
described elsewhere
\citep[cf.][]{borucki:2010a,koch:2010,batalha:2010,caldwell:2010,gilliland:2010,jenkins:2010a,jenkins:2010b}.

\citet[hereafter Paper I]{prsa:2011} catalogs 1879 eclipsing and ellipsoidal binary systems identified
in the first \Kepler\ data release \citep{borucki:2010b}.  The catalog lists
the \Kepler\ ID, ephemeris, morphological type, physical parameters and
third-light contamination levels from the \Kepler\ Input Catalog, and
principal parameters determined by a neural network analysis of the phased
light-curves.
For the detached and semi-detached binaries the computed
principal parameters are
the ratio of the temperatures $T_2/T_1$, 
the sum of the fractional radii $\rho_1+\rho_2\equiv(R_1+R_2)/a$, where $a$ is the semi-major axis of the orbit, 
the radial and tangential components of the eccentricity $e\sin\omega$ and $e\cos\omega$, respectively, where $\omega$
is the argument of periastron,
and the sine of the inclination, $\sin i$.
For the over-contact systems the computed parameters are $T_2/T_1$,
the photometric mass ratio $q_{\mathrm ph}$,
the fill-out factor
$F=(\Omega-\Omega_{\mathrm L_2})/(\Omega_{\mathrm L_1}-\Omega_{\mathrm L_2})$
where $\Omega$ is the surface potential \citep{wilson:1979},
and $\sin i$.
An online
version of the catalog also provides phased and un-phased light curves for all
the systems (\anchor{http://keplerEBs.villanova.edu}{http://keplerEBs.villanova.edu}).

With the second \Kepler\ data release we are updating the catalog in several ways:
\begin{enumerate}

\item
The light curves of Kepler Objects of Interest (KOI's) flagged as possibly
containing planetary transit events and subsequently rejected as planet
transits have been examined.
If these are identified as eclipsing binaries, or as blends containing an EB,
we computed their ephemerides and
included them in this catalog (\S\ref{sec:additions}).

\item
There are 77 systems identified earlier but with only single events in the
first data release.  Periods for these can now be determined and they are part
of the catalog.  124 additional systems were identified as EBs in the Q2 data
from the \Kepler\ Transit Planet Search (TPS) output.  Another 19 eclipsing
binary systems were not in the first data release for proprietary reasons have
now also been included (\S\ref{sec:additions}).

\item
42 objects cataloged in Paper I as EBs have since been re-classified as
short-period pulsating variables and these have been removed
(\S\ref{sec:deletions}).

\item
An analysis of flux variations of individual pixels within a \Kepler\ target
aperture has revealed that 58 of the identified eclipsing binaries are blended
objects where the eclipsing system is not the \Kepler\ target star.  These EBs
are not centered in the target aperture and have been removed pending
re-observation with a re-centered aperture (\S\ref{sec:blended});

\item
All ephemerides have been recomputed.  The baseline has increased
substantially, from 34 days to 125 days, resulting in an increase in precision.
(\S\ref{sec:ephemerides}).
\end{enumerate}

In addition, we point out 10 systems that show evidence for the presence of
third-bodies.  Four of these have extra transit features in their light
curves (tertiary eclipses) (\S\ref{sec:tertiarySystems}) and 8, including 2 of
the tertiary eclipsing systems, show large eclipse timing variations
(\S\ref{sec:etvs}).

\section{Catalog Updates}

The initial release of the catalog featured 1879 unique objects that contained
the signature of an eclipsing binary and/or ellipsoidal variable in the first
\Kepler\ data release (Q0+Q1). In this update, the following data sources were
used to add or remove objects from the catalog.

\subsection{Catalog Additions\label{sec:additions}}

The catalog of Kepler Objects of Interest
\citep[KOI;][]{borucki:2010b,borucki:2011} lists all detected planets and
planet candidates. There is an inevitable overlap between planet transits and
severely diluted binaries or binaries with low mass secondaries. As part of
the main Kepler effort, these targets are vetted for any EB-like signature,
such as depth change of successive eclipses (the so-called \emph{even-odd}
culling), detection of a secondary eclipse that is deeper than what would be
expected for a $R < 2R_\mathrm{Jup}$ planet transit (\emph{occultation}
culling), \ion{He}{0}-core white dwarf transits (\citealt{rowe:2010}; \emph{white dwarf}
culling), and spectroscopic follow-up where large amplitudes or double-lined
spectra are detected (\emph{follow-up} culling).
High resolution direct imaging (AO and speckle) and photo-center centroid shifts
also indicate the presence of background EBs. 
The culling criteria and the
results are presented in detail by \citet{borucki:2011}.  292 of these, tagged
with \verb|KOI|, are now in the main catalog.

The output of the Transit Planet Search \citep{jenkins:2010c} provides
transit event detection statistics for each light curve.
The Single Event Statistic (SES) is the maximum detection statistic found for a light curve.
The Multiple Event Statistic (MES) is the maximum detection statistic after folding the data with different periods.
As periods up to the length of the data are considered, strong single transits or eclipses are detected as well as
series of transits \citep[Eq.~11]{jenkins:2002}.
The total number of events exceeding the detection threshold (Threshold Crossing Events, TCEs)
in the Q2 data is over 86,000.
Of those, most are data anomalies.  To pick the
most suitable EB candidates from the list of all TCEs, we selected those for
which the MES-to-SES ratio is larger than $\sqrt2$ to reflect a detection of
$\geq2$ events in the time series.
This filtering yielded $\sim5000$
candidate TCEs that were cross-checked against already cataloged EBs and KOIs.
Those, as well as all duplicate entries, were removed and the final list of
candidates contained 2153 targets.  We manually checked all of them and found
124 new EBs.  These targets are flagged with \verb|NEW| in the main catalog.

The initial catalog contained 101 EBs with single events, objects with periods
longer than the Q1 time span, or objects for which we were unable to determine the periods
from Q1 data alone  (we required 2 eclipses be visible).  In this update we
provide the ephemerides for 77 of these EBs, with 24 EBs still remaining
uncertain because of the periods longer than the Q1+Q2 time span of $\sim130$
days.

At the time of Q1 data release in June 2010, 19 eclipsing binaries were held
back for Guest Observer (GO) programs. Q1 data were made public in December 2010 and Q2 data
are being released now. These targets are tagged with \verb|Q1HB| in the main
catalog.

Two EBs whose light curves contained anomalous eclipse events during Q1 were
not in the first data release.  Both of these have now been added to the
catalog and are further discussed below in \S\ref{sec:tertiarySystems}.

\subsection{Catalog Deletions\label{sec:deletions}}

The time span of Q1 data did not allow for reliable detections of period
drifts that would be typical of pulsating single stars and would be atypical
of binaries.  With the added Q2 data, we ran a cross-check against the short
period pulsators presented in \citet{debosscher:2011}.  That check yielded 42
objects which, after manual inspection of their light curves,
were subsequently removed from the EB catalog.
We list
the \Kepler\ Identification numbers (KID) for these objects along with their
period of variability in Table \ref{tab:puls}.

Other cross-checks have been performed, namely against the list of
chromospherically active stars \citep{basri:2011}, \citet{coughlin:2010}'s
list of low mass binaries, and a GO-reported list \citep{morrison:2011}, but
neither new EB targets nor any conclusive non-EB stars have been found.

\begin{deluxetable}{cc|cc|cc}
\tablecaption{Short period pulsating stars culled from catalog\label{tab:puls}}
\tablewidth{0pt}
\tablehead{ \colhead{KID} & \colhead{$P_0$ [d]}   & \colhead{KID} & \colhead{$P_0$ [d]}  & \colhead{KID} & \colhead{$P_0$ [d]}}
\startdata
 1849235 & 0.3192  &  4544967 & 0.12397 &  6032172 & 0.07045 \\
 2168333 & 0.0921  &  4569150 & 0.20622 &  6231538 & 0.16292 \\
 3338680 & 0.17131 &  4577647 & 0.21684 &  6606229 & 0.31166 \\
 3424493 & 0.73872 &  4940217 & 0.37873 &  6963490 & 0.28008 \\
 3648131 & 0.12974 &  5108514 & 0.28557 &  7300184 & 0.1715 \\
 3965879 & 0.30592 &  5358323 & 0.155   &  7900367 & 0.15098 \\  
 4072890 & 0.29873 &  5900260 & 0.11537 &  7915515 & 0.13216 \\ 
 \\
 8264404 & 0.21285 &  9051991 & 0.19253 &  9851822 & 0.13648 \\
 8330102 & 0.11537 &  9306095 & 0.19658 & 10350769 & 0.63719 \\
 8453431 & 0.1443  &  9368220 & 0.37087 & 10355055 & 0.09056 \\
 8493159 & 0.27303 &  9368524 & 0.19034 & 10415087 & 0.27666 \\
 8585472 & 0.1595  &  9649801 & 0.1388  & 11027806 & 0.37676 \\
 8845312 & 0.31945 &  9716523 & 0.9217  & 11769929 & 0.19774 \\
 9050337 & 0.11394 &  9773512 & 0.21719 & 12216817 & 0.24601 \\
\enddata
\end{deluxetable}

\subsection{Re-identifications within Blended Sources\label{sec:blended}}

The first \Kepler\ EB catalog contained a small fraction of blends,
cases where the eclipse signature is from a nearby source in the photometric
aperture.
Although there is variation across the field-of-view,
on average 47\% of the energy from a star centered on a pixel
falls within that pixel
and the photometric response function has a typical 
95\% encircled energy diameter of 4 pixels
\citep{bryson:2010}.
Since each pixel is 4\arcsec\ across, blending of sources is expected.
In constructing the original catalog, obvious blends were
identified and removed and/or reassigned to the appropriate point source.  We
build upon this work by performing pixel-level tests that pinpoint the blended
cases and identify the correct EB sources.  These tests, summarized here, are
similar to those used to identify false positives amongst the \Kepler\ exoplanet
candidates and are described in detail by \citet{brysonetal:2011}.

\begin{figure}
\plotone{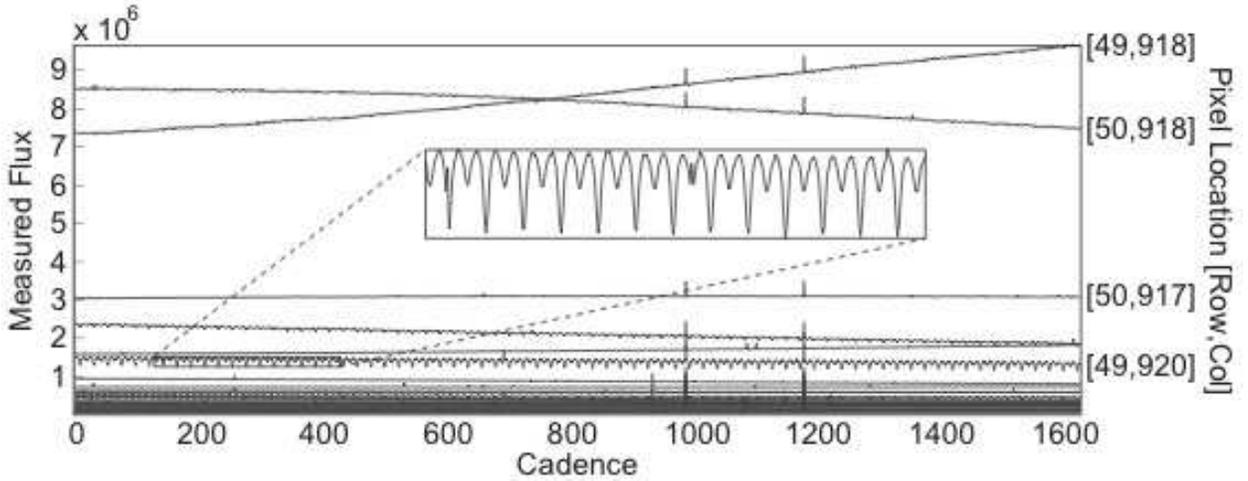}
\caption{Flux-time series for KID\,5041847. The time series for the pixel in
  which the target falls upon, pixel [50,918], has no
  observable eclipse signature; however, the time series for pixel [49,920],
  enlarged in the pop-out, reveals a clear eclipse signature.
  The target's pixel coordinates originate from a corner of the appropriate CCD module.
\label{fluxtime}}
\end{figure}

\begin{figure}
\plotone{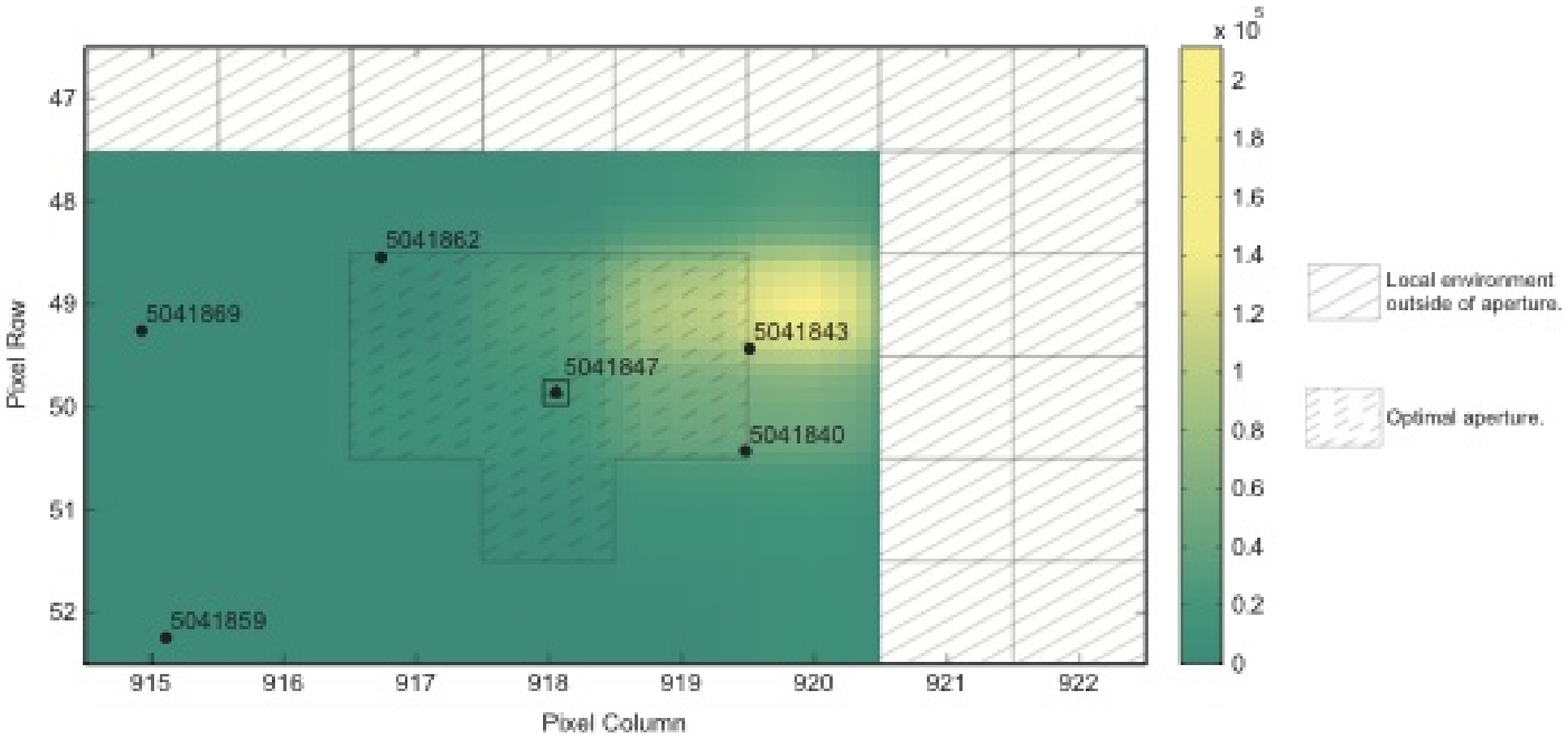}
\caption{An expanded difference image for target KID\,5041847 showing a greater view
of the local environment including an overlay of the target's optimal aperture.
This image clearly shows that the pixel with the greatest in-eclipse versus out-of-eclipse
difference in flux is not centered on KID\,5041847. The true eclipsing binary star is KID\,5041843. 
\label{difference}}
\end{figure}

The probable blends are pinpointed by an automated analysis of each target's
photometric aperture. For each pixel within an aperture, the relative depth of
the transit observed in the Q1 flux-time series is calculated using averaged
in-transit times and averaged times just before and after the eclipses. A
target is flagged as a probable blend when the deepest eclipse occurs on a
pixel adjacent to the pixel that the target source falls upon. Once a target
is flagged as a probable blend, a manual inspection of the aperture
flux-time series and difference image is conducted to validate the blend, and
when confirmed, to identify the correct source. The pixel-time series, shown in
Fig.~\ref{fluxtime}, is an example of a typical blend scenario. The time
series for the pixel that the target EB falls upon shows no eclipse signature,
whereas the time series for an adjacent pixel is showing a clear EB
signature. The photometric location of the transit is revealed by the
aperture difference image, Fig.~\ref{difference}. The difference image is
created by subtracting each pixel's averaged in-eclipse values with its
averaged out-of-eclipse values, and in a typical case like the one presented in Fig.~\ref{difference},
the exact location of the EB becomes apparent. An overlay of the target's stellar
environment, acquired from the \Kepler\ Input Catalog, shows the location of
the sources surrounding the target and clearly identifies the correct EB
source.

After thorough inspection, 58 of the objects included in the first catalog (Paper I) were found to be blended
systems where the target star is not itself an EB.
The corrected KIDs for these systems are listed in Table \ref{correctedKIDs}.
These objects have been re-targeted, that is re-centered in optimal apertures, for observation starting in Q8
and have been removed from the catalog pending further observations.
Of the former KOIs examined as potential EBs for inclusion into the catalog (\S\ref{sec:additions}),
172 were found to be blended light curves where the EB component was not the target star.
The eclipsing binaries in the blends have been similarly identified and will be added to the target list
for observation starting with Q10.

Two overlapping EB systems were discovered during this analysis; both were uncovered as a second EB
signature within a target EBÕs light curve.  KID\,3437778 and KID\,5983351 are
the new EBs and their eclipses can be seen superposed in the flux-time series of KID\,3437800 and KID\,5983348,
respectively.

\begin{deluxetable}{cc|cc|cc}
\tablecaption{Corrected \Kepler\ IDs of Blended EBs\label{correctedKIDs}}
\tabletypesize{\footnotesize}
\tablewidth{0pt}
\tablehead{\colhead{Original KID} & \colhead{Corrected KID} & \colhead{Original KID} & \colhead{Corrected KID}
  & \colhead{Original KID} & \colhead{Corrected KID}}
\startdata
  7432476 &   7432479 &   5649837 &   5649836 &   5392871 &   5392897 \\
  6470521 &   6470516 &  10491544 &  10491554 &   9075708 &   9075704 \\
  3338674 &   3338660 &   7590723 &   7590728 &   5041847 &   5041843 \\
  3735634 &   3735629 &   7707736 &   7707742 &   4073730 &   4073707 \\
  9535881 &   9535880 &   9935242 &   9935245 &  11825056 &  11825057 \\
  3549993 &   3549994 &  11247377 &  11247386 &   4579313 &   4579321 \\
  9851126 &   9851142 &   8780959 &   8780968 &   5816811 &   5816806 \\
 10095484 &  10095469 &   8263752 &   8263746 &  10743597 &  10743600 \\
  5467126 &   5467113 &   8589731 &   8589754 &   9456932 &   9456933 \\
  6182846 &   6182849 &   8620565 &   8620561 &   6233890 &   6233903 \\
  9834257 &   9773869 &   7691547 &   7691553 &   5730389 &   5730394 \\
  7376490 &   7376500 &   6233483 &   6233466 &   9366989 &   9366988 \\
  5956787 &   5956776 &   9468382 &   9468384 &   5020044 &   5020034 \\
  4474645 &   4474637 &   6286155 &   6286161 &   8647295 &   8647309 \\
  8097897 &   8097902 &   6314185 &   6314173 &   6312534 &   6312521 \\
  2451721 &   2451727 &   5560830 &   5560831 &  10747439 &  10747445 \\
  9664387 &   9664382 &   6677267 &   6677264 &   3335813 &   3335816 \\
  7516354 &   7516345 &   5390342 &   5390351 &   3446451 &   3547315 \\
  6058896 &   6058875 &   5565497 &   5565486 & & \\
  5022916 &   5022917 &   7910148 &   7910146 & & \\
\enddata
\end{deluxetable}

\subsection{Updating the Ephemerides\label{sec:ephemerides}}

With the inclusion of Q2 data, the duration of the light curves increases by a
factor of 3.7, so an update of the ephemerides was appropriate. We improved
the ephemerides in two steps: First, using a software tool, {\tt kalahari}, that overlays a cursor
cross-hair on the light curve, we selected and centered by eye the first and
last occurring clean eclipses in the Q0 through Q2 light curves.
This was
greatly facilitated by using the previous ephemerides published in Paper I to
predict the first and last eclipse; in the vast majority of cases the
ephemeris was accurate enough to clearly identify both eclipses and an
unambiguous cycle count. These two eclipses were then used to compute a more
precise period.  An assortment of median, linear, and cubic polynomial
detrending options were used to allow combining the different quarters (and
discontinuous sections of Q2) together.  Phase-folded figures were
automatically generated for every system and checked for correctness of the
ephemeris---even slight errors in the ephemeris were readily
apparent in the phase folded light curves.
We estimate the
uncertainties in this {\tt kalahari} manual eclipse selection method to be
roughly 50--700\,s in the initial epoch of eclipse center, T0, and 0.2--90\,s in the period, P,
with the shorter period systems
giving the higher precision.  This ``hands on'' approach allowed us to
visually inspect every light curve and use judgment in the selection of T0, a
task that is typically problematic for automated methods.
We chose the barycentric Julian day (BJD) of the first good eclipse
to define the epoch of eclipse center.
This is mainly for convenience.
The epoch will not change as more data are added, it places
cycle number zero at the start of the light curve, and it makes checking the
epoch by users of the catalog relatively easy.
In addition, the times are now in BJD,
obtained directly from the MAST {FITS} files,
and are not approximated as in Paper I.

Kepler experienced one Safe Mode event and four spacecraft attitude tweaks in
Q2 \citep[Table~5]{dr7}, each creating discontinuities in the light curves of various amplitude.
In many cases these discontinuities are obvious.
However, we caution that in noisy,
shallow-eclipse cases, the discontinuity near BJD 2455079.18 is sometimes flagged as
an eclipse event triggering a false detection of a periodicity.
Similarly, portions of the light curve contain a low amplitude modulation that
can sometimes mimic a periodicity.  Periods very near $3\fd00$ and $0\fd133$
should be treated skeptically, as these may be instrumental in origin
(reaction wheel momentum desaturation cycle and focus change due to reaction
wheel housing heaters).  See the {\it Kepler Data Characteristics Handbook}
\citep{christiansen:2011} for these and other important data issues.

The second step of ephemerides determination refines the above estimates by
using them as inputs to the {\tt ebai} engine.
Part of the EBAI project (Eclipsing Binaries via Artificial Intelligence),
the {\tt ebai} engine is a back-propagating neural network trained on synthetic eclipsing binary data
that is able to quickly determine principal parameters for large numbers of observed light curves \citep{prsa:2008}.
Its performance on EBs in the first \kepler\ data release is described in Paper I.
We adopted the {\tt ebai} estimates of T0 as the final values for the ephemerides.

\subsection{Catalog Description}

The updated catalog contains 2165 eclipsing binaries.
Each EB is identified by its \Kepler\ ID in column 1.
Its ephemeris, in days,  is given in Columns 2 \& 3 ($\mathrm{BJD}_0$, and $P_0$) and
subsequent columns contain: morphological classification (Column 4) as one of \D\ (detached),
\SD\ (semi-detached), \OC\ (overcontact) , \ELV\ (ellipsoidal variable) or \UNC\ (unclassified);
the source of the target (Column 5) which tracks the origin of the added target:
\verb|CAT| if it appeared in the first catalog release, \verb|Q1HB| if it was
held back at the time of the initial release but is now public, \verb|KOI| if
it is a rejected Kepler Object of Interest due to the detected EB signature,
and \verb|NEW| if it was a newly discovered EB;
the systems \Kepler\ magnitude (Column 6);
and input catalog parameters, $T_\mathrm{eff}$ in K (Column 7),
$\log\mathrm{g}$ in cgs units (Column 8), $E(\bv)$ (Column 9), and the estimated contamination (Column 10);
the principal parameters: $T_1/T_2$ (Column 11), the scaled sum of the radii $\rho_1+\rho_2\equiv(R_1+R_2)/a$ (Column 12),
the fillout factor $F$ (Column 13),
the radial and tangential components of eccentricity $e\sin\omega$ and $e\cos\omega$ (Columns 14 \& 15),
the mass ratio $q$ (Column 16), and the sine of the inclination $\sin i$ (Column 17).

\begin{deluxetable}{ccccccccccccccccc}
\tablewidth{0pt}
\tabletypesize{\scriptsize}
\tablecaption{Catalog of EBs in \Kepler\ Q0--2 Data\label{tab:example}}
\tablehead{
  \colhead{KID} & \colhead{$\mathrm{BJD}_0$} & \colhead{$P_0$ [days]} & \colhead{Type} & \colhead{Source} &
  \colhead{$K\mathrm{mag}$} & \colhead{$T_\mathrm{eff}$[K]} & \colhead{$\log\mathrm{g}$ [cgs]}
  		& \colhead{$E(\bv)$} & \colhead{contam} \\
   & & \colhead{$T_2/T_1$} & \colhead{$\rho_1+\rho_2$}  & \colhead{Fillout} & \colhead{$e\sin\omega$}
   & \colhead{$e\cos\omega$} & \colhead{$q$} & \colhead{$\sin i$}
}
\startdata
01026032.00 & 54966.773843 &  8.460438 &  \D & \CAT & 14.813 &  5715 & 4.819 & 0.107 & 0.266 \\
& & 0.85956
& 0.12451 & \nodata &  0.05515 &  0.01308 & \nodata & 0.99687 \\
01026957.00 & 54956.011753 & 21.762784 &  \D & \KOI & 12.559 &  4845 & 4.577 & 0.036 & 0.034 \\
& & 0.49053
& 0.18848 & \nodata & -0.06237 & -0.07830 & \nodata & 0.98538 \\
01433962.00 & 54965.325203 &  1.592691 &  \D & \KOI & 15.470 &  4349 & 4.634 & 0.067 & 0.609 \\
& & 0.78423
& 0.11622 & \nodata & -0.12883 &  0.07820 & \nodata & 0.99716 \\
01571511.00 & 54954.506187 & 14.021624 &  \D & \KOI & 13.424 &  5804 & 4.406 & 0.101 & 0.011 \\
& & 0.82928
& 0.13522 & \nodata & -0.10259 & -0.02367 & \nodata & 0.99416 \\
01725193.00 & 55005.663605 &  5.926658 &  \D & \NEW & 14.502 &  5802 & 4.384 & 0.146 & 0.772 \\
& & 0.82976
& 0.23817 & \nodata & -0.01015 &  0.05277 & \nodata & 0.97561 \\
01996679.00 & 54979.068748 & 20.000276 &  \D & \KOI & 13.884 &  5914 & 4.334 & 0.119 & 0.018 \\
& & 0.69915
& 0.13037 & \nodata & -0.14132 &  0.16568 & \nodata & 0.99437 \\
02010607.00 & 54974.583000 & 18.627229 &  \D & \CAT & 11.347 &  6122 & 4.344 & 0.056 & 0.065 \\
& & 0.76403
& 0.12954 & \nodata & -0.11481 &  0.03654 & \nodata & 0.99506 \\
02162635.00 & 55009.129448 & \nodata      &  \D & \KOI & 13.862 &  4787 & 3.567 & 0.160 & 0.061 \\
& & \nodata
& \nodata  & \nodata &  \nodata  &  \nodata  & \nodata &  \nodata \\
02162994.00 & 54965.631839 &  4.101588 &  \D & \CAT & 14.162 &  5410 & 4.532 & 0.099 & 0.189 \\
& & 0.86621
& 0.18990 & \nodata & -0.06236 &  0.00060 & \nodata & 0.99798 \\
02305372.00 & 54965.963928 &  1.404636 &  \D & \CAT & 13.821 &  5664 & 3.974 & 0.158 & 0.267 \\
& & 0.51753
& 0.59250 & \nodata & -0.00898 & -0.00365 & \nodata & 1.00256 \\
02305543.00 & 55003.400185 &  1.362339 &  \D & \NEW & 12.545 &  5623 & 4.486 & 0.064 & 0.001 \\ 
& & 0.87172
& 0.31464 & \nodata & -0.03268 &  0.01347 & \nodata & 0.97667 \\
02306740.00 & 54987.038258 & 10.307175 &  \D & \CAT & 13.545 &  5647 & 4.228 & 0.117 & 0.241 \\
& & 0.86225
& 0.13727 & \nodata & -0.11561 &  0.02030 & \nodata & 0.99919 \\
02308957.00 & 54965.169838 &  2.219736 &  \D & \CAT & 14.520 &  5697 & 4.343 & 0.148 & 0.649 \\
& & 0.95548
& 0.46031 & \nodata & -0.00715 &  0.01505 & \nodata & 0.98274 \\
\enddata
\tablecomments{Table \ref{tab:example} is published in its entirety in the electronic edition of the Astronomical Journal.
A portion is shown here for guidance regarding its form and content.}
\end{deluxetable}

\section{Interesting Objects in the Catalog}

\subsection{Tertiary Eclipses\label{sec:tertiarySystems}}

\begin{figure}
\epsscale{0.9}
\plotone{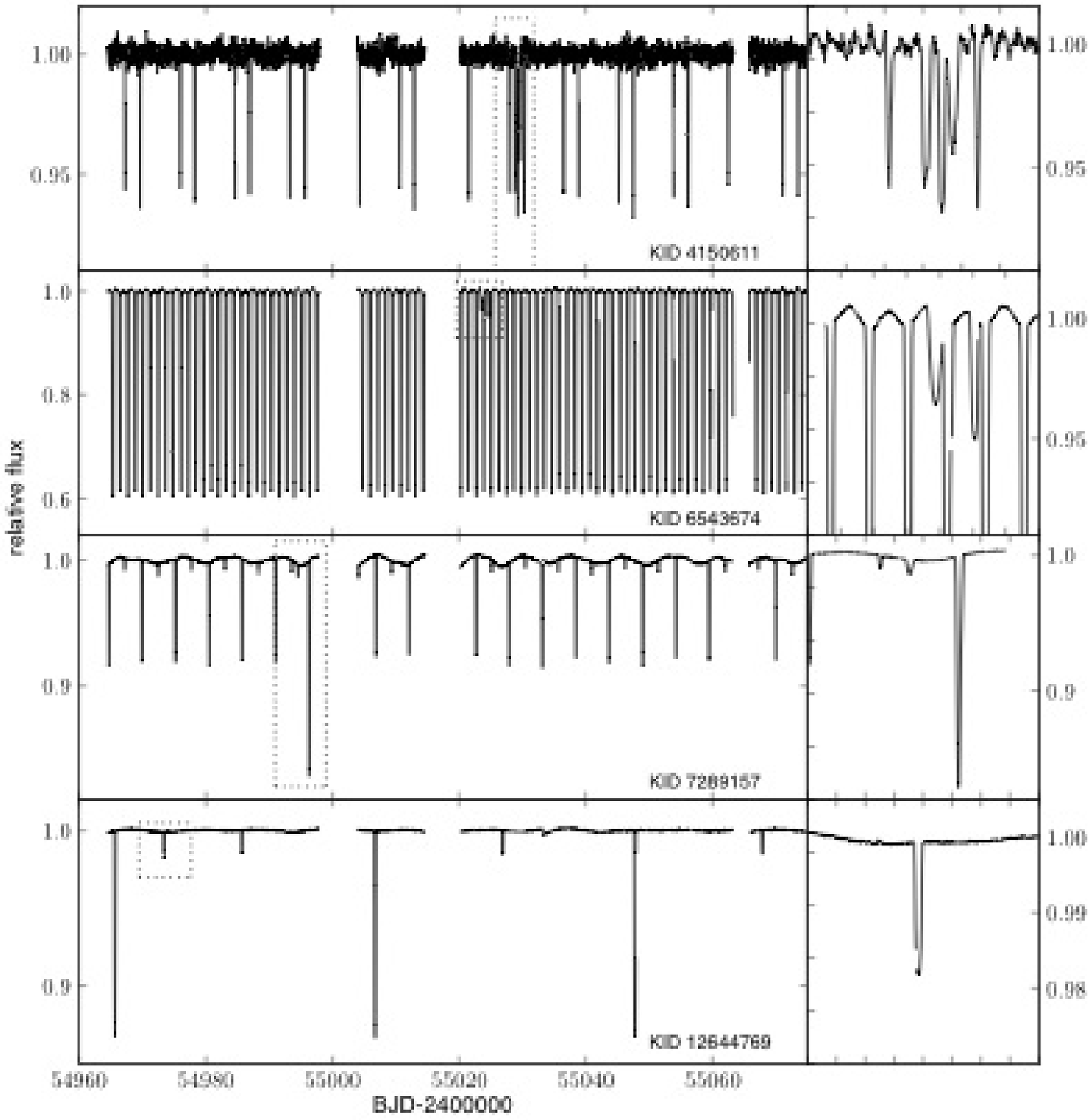}
\caption{Four EBs with significant tertiary eclipse events.  The systems are (top to
  bottom): KID\,4150611 ($P=8\fd65308$), 6543674 ($P=2\fd39105$),
  7289157 ($P=5\fd26627$), and 12644769 ($P=41\fd0781$).
  The Q1+Q2
  light curve for each system is shown on the left with the tertiary event marked with
  a dashed box.  On the right is an expanded region
  around the anomalous event.}  
\label{tertiaryEclipses}
\end{figure}

The search for circumbinary planets in the \kepler\ data includes looking for
transits with multiple components \citep[e.g.][]{deeg:1998,doyle:2000}. 
Transit patterns with multiple components are caused by a slowly moving planet
crossing in front of the eclipsing binary;
it is alternately silhouetted by the motion of the
background binary stars as they orbit about each other.
Circumbinary transits
can thus produce predictable but non-periodic features of various shapes and
depths.  We have been looking for such features in the Catalog and, 
in the process, have identified several tertiary
eclipses where the depth of each event is, in most cases, too deep to be the
transit of a planet but is, instead, an eclipse by a third (sub-)stellar
body.
In Fig.~\ref{tertiaryEclipses} we show 4 such systems illustrating the variety of "transit signatures"
that can be produced and discuss each system in turn in what follows.

KID\,4150611 ($P=8\fd65308$) shows a series of grazing eclipses.
The system is bright, {\it Kep}$_m=7\fm889$, yet very ``noisy'' with a \Kepler\ Input Catalog (KIC) temperature,
$T_\mathrm{eff}=6623$\,K suggesting a mid-F type star undergoing $\delta$\,Sct oscillations.
A triplet event can be seen between the primary and secondary eclipses near the
middle of the light curve which is somewhat puzzling.
The short, $0\fd7$ duration of the triple event occurs while the EB is out-of-eclipse
so it cannot be a single third-body transiting the EB as that would result in only two dips.
A more plausible model has a short-period
binary system transiting one of the EB components
similar to the KID\,5897826 system discussed below.
That is a quadruple system consisting of two binary systems where one of the systems is eclipsing.
Another possibility we are considering is that the light curve is the composite of an hierarchical triple,
the F-star and the short-period binary, plus
an additional eclipsing binary either physically associated with the triple or simply a blend.

KID\,6543674 ($P=2\fd390105$) is a shorter period EB with deep eclipses from two nearly equal components seen close to edge-on.
Their separation is small enough that mutual tidal forces are distorting the stars yielding
the distinct out-of-eclipse ellipsoidal (aspect) variations
which are readily seen in the expanded box on the right.
There are two tertiary eclipses separated by $1\fd2$ which is
consistent with the model of a single third-body passing in front of the
EB and being alternately silhouetted by the EB components as they orbit one another.
This system also has eclipse timing variations (\S\ref{sec:etvs}, Fig.~\ref{newupdatefig01a}) that
may arise from the light time effect as the third-body orbits that binary.

KID\,7289157 ($P=5\fd26627$) has two tertiary events less than
two days apart with the second event coinciding with a primary eclipse.
The depth of a transit in a binary system is shallower than in the equivalent 
single star case as flux from the non-transited binary component diluting
the transit signature.
When a transit occurs during an eclipse, the dilution is significantly reduced
yielding a deeper transit dip as has happened here.
Like HD\,6543679, this system also has eclipse timing variations (Fig.~\ref{newupdatefig01b}) 
but with a significantly larger amplitude and large a discrepancy between primary and secondary times.
Dynamical interactions will need to be considered to understand this system.

KID\,12644769 ($P=41\fd0781$) has a single extra event in the light curve during Q1
with a depth of slightly less that $\la2\%$.
With only a single event, one cannot rule out a blend with a long period background EB
and we do note that this feature is slightly deeper than the secondary eclipses.
The event is potentially interesting considering the photometrically derived stellar
parameters for this system in the KIC suggests that the components are late-K or M-dwarfs.
($T_\mathrm{eff}=4051$\,K, $\log g=4.48$, radius$=0.74$).
If so, they imply that the radius of the transiting body is $\la2 R_\mathrm{Jup}$ and that
there may be a sub-stellar object orbiting this system.

These and other tertiary events are being studied and will be further described
in an upcoming paper \citep{doyle:2011}.

\begin{deluxetable}{cccc}
\tablecaption{Significant Tertiary Eclipse Events during Q1 and Q2}
\tablewidth{0pt}
\tablehead{\colhead{\kepler\ ID} & \colhead{Event} &
\colhead{Mid-time} & \colhead{Mid-event depth} \\
	\colhead{} & \colhead{(Fig.\ref{tertiaryEclipses})} & \colhead{[2400000-BJD]} & \colhead{[mag]}} 
\startdata
\phn4150611 & 1 & 55028.9 & 0.0630 \\
                  & 2 & 55029.3 & 0.0745 \\
                  & 3 & 55029.6 & 0.0502 \\                 
\phn6543674 & 1 & 55023.5 & 0.0500 \\
                  & 2 & 55024.7 & 0.0644 \\
\phn7289157 & 1 & 54994.6 & 0.0105 \\
12644769 & 1 & 54973.4 & 0.0204 \\
\enddata
\end{deluxetable}

Finally, another eclipsing binary with tertiary eclipses, KID\,5897826 (KOI-126), was described in detail by \citet{carter:2011}. 
It consists of two M-dwarfs in a 1\fd77 mutual orbit, which itself orbits around a slightly evolved $1.35\,M_\odot$ primary star with a 33\fd9 orbit. 
Each time the M-dwarfs pass in front of the primary, they give rise to two $\sim1.5$\% deep, 
transit-shaped events, which are often superimposed due to the relative phasing of the small and large orbits, and distorted due to the
acceleration of the M-dwarfs by each other during the transit across the primary.
Eclipses between the M-dwarfs are also seen at the beginning of the dataset but their depths are reduced to zero as the primary
causes their orbits to precess into a non-eclipsing inclination.
Besides being a dramatic demonstration of dynamical interactions in a triple-star system, dynamical fits produced a measurement of masses
($0.2413\pm0.0030\,M_\odot$ and $0.2127\pm0.0026\,M_\odot$) and radii ($0.2543\pm0.0014\,R_\odot$ and
$0.2318\pm0.0013\,R_\odot$) for two M-dwarfs, a valuable test of theoretical stellar structure models.

\subsection{Eclipse Timing Variations\label{sec:etvs}}

\begin{deluxetable}{cllcc}
\tablecaption{EBs with eclipse timing variations\label{tab:oc}}
\tabletypesize{\scriptsize} \tablewidth{0pt} \tablehead{ \colhead{KID} &
  \colhead{$P_{\rm prim}$} & \colhead{$P_{\rm sec}$} & \colhead{O-C range,
    primary} & \colhead{O-C range, secondary} \\ \colhead{} & \colhead{(d)} &
  \colhead{(d)} & \colhead{(min)} & \colhead{(min)} }
\startdata
\phn5771589 & $10.7379\phn\phn\phn\phn  \pm 0.0023$      & $10.7374\phn\phn\phn\phn\pm 0.0014$              & 140 & 89  \\
\phn6543674 & $\phn2.39104706                   \pm 0.000003$ & $\phn2.39104110 \pm 0.000006$                          & 2.4  & 1.2 \\
\phn6545018 & $\phn3.99144 \phn\phn\phn \pm 0.00008$    & $\phn3.99143 \phn\phn\phn\pm 0.00009$           &  17  & 18  \\
\phn7289157 & $\phn5.26585 \phn\phn\phn \pm 0.00001$    & $\phn5.26625 \phn\phn\phn\pm 0.00010$           &  1.9 & 15  \\
\phn7668648 & $27.7828 \phn\phn\phn\phn \pm 0.0050$       & $27.7976 \phn\phn\phn\phn\pm 0.0005$             &  35  & 2.7 \\
\phn7955301 & $15.3225 \phn\phn\phn\phn \pm 0.0047$       & $15.3196 \phn\phn\phn\phn \pm 0.0045$            & 137 & 169 \\
\phn9714358 & $\phn6.4739 \phn\phn\phn\phn \pm 0.0004$ & $\phn6.4738 \phn\phn\phn\phn\pm 0.0003$        &  42  & 41  \\
      10319590 & $21.3273 \phn\phn\phn\phn \pm 0.0013$      & $21.3227 \phn\phn\phn\phn \pm 0.0009$             &  20  & 8   \\
\enddata
\end{deluxetable}

In an EB, one normally expects the primary eclipses to be uniformly spaced in
time.  However, mass transfer from one star to the other or the presence of a
third star in the system can give rise to changes in the orbital period, which
in turn will change the time interval between consecutive eclipse events.  The
eclipse times will no longer be described by a simple linear ephemeris, and
the deviations (usually shown in the ``O-C''
diagram) will contain important clues as to the origin of the period change.
We have begun to systematically measure the times of primary and secondary
eclipse for the {\em Kepler} sample of EBs classified as detached (D) and
semidetached (SD).  This is a difficult task, owing to
a host of intrinsic variabilities and systematic problems.
These include large spot modulations that may or may not be in phase with
the eclipses, pulsations and/or noise in the out-of-eclipse regions, thermal
events and cosmic ray hits that make the normalization of the light curves
hard to automate, and eclipses falling partially or completely in data gaps.
This work will be fully described in an upcoming paper \citep{orosz:2011}. 
We present here some
interesting cases of EBs with O-C variations evident in the Q0--Q2 data.

Briefly, these basic steps are followed to measure the times of eclipse.  (i)
The light curve is detrended quarter by quarter and combined. (ii) The
ephemeris is determined, and (iii) the light curve is phased based
on the ephemeris values.
(iv) A simple function of the form $y=x^n$, where $n$ is not necessarily an
integer, is fit to half of the eclipse profile and then reflected to the other
half.  Finally, (v) the light curve is unfolded, and the mean profile is fit to
individual events after an additional local de-trending is applied to obtain
the eclipse time.  In some cases, an automated code was able to set the limits
of the local fitting on its own, and in other cases, it was necessary to
specify the fitting limits manually.  A linear ephemeris is fitted to the
times, and the O-C diagram is generated.  The typical uncertainties in the
individual times range from about 30 seconds in the best cases to around two
minutes for cases with noisy out-of-eclipse regions (where the ``noise'' can
be due to spot modulations or pulsations in addition to shot noise).

\begin{figure}
\epsscale{0.8}
\plotone{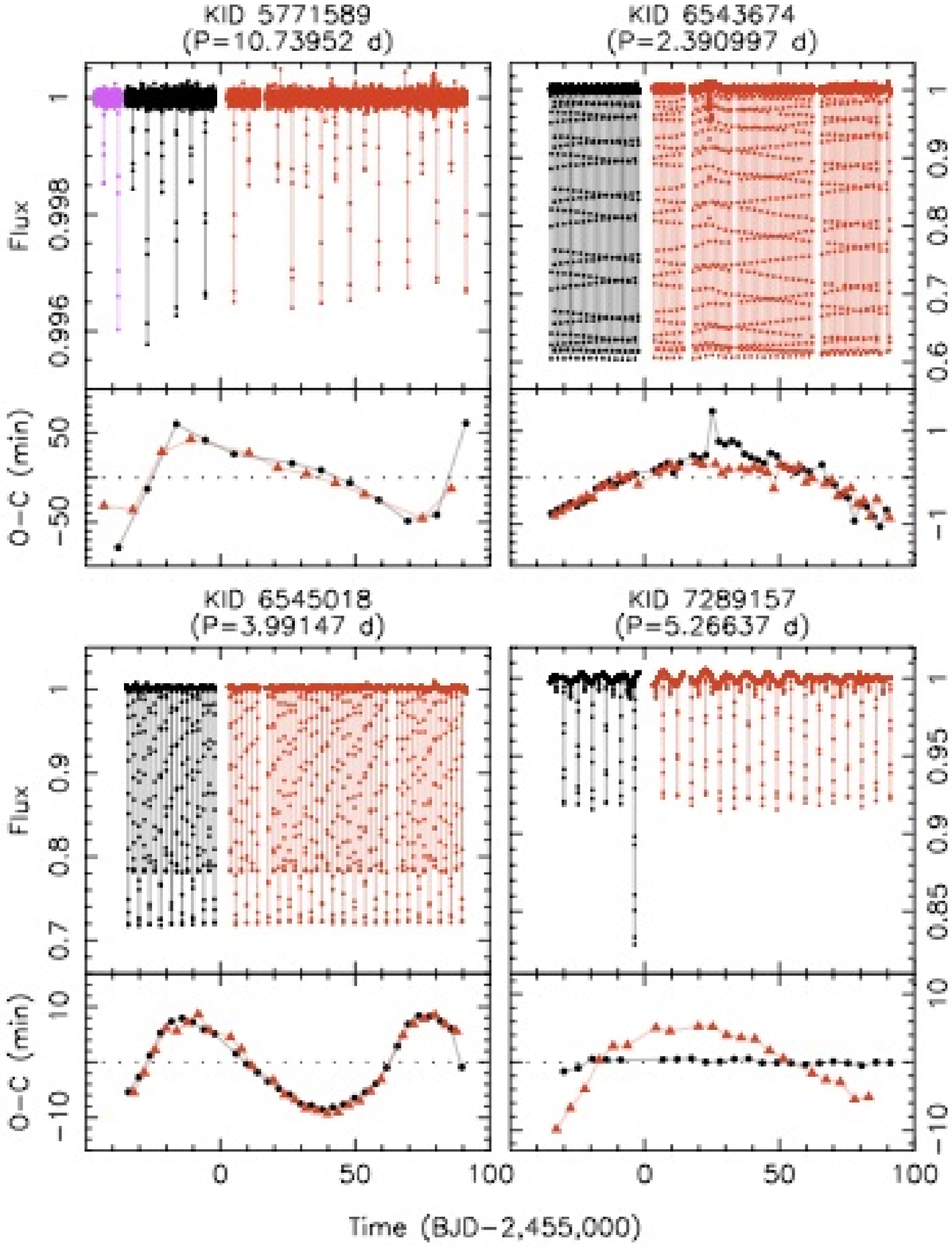}
\caption{EBs with a significant signal in the O-C diagram.
The normalized light curves are shown in the upper parts of each panel
where the different colors correspond to data from different quarters,
and the O-C diagram with curves for the primary (filled circles) and
secondary (filled triangles) eclipses are shown in the lower parts of
each panel.  KID\,6543674 and KID\,7289157 have tertiary eclipses.
\label{newupdatefig01a}}
\end{figure}

\begin{figure}
\plotone{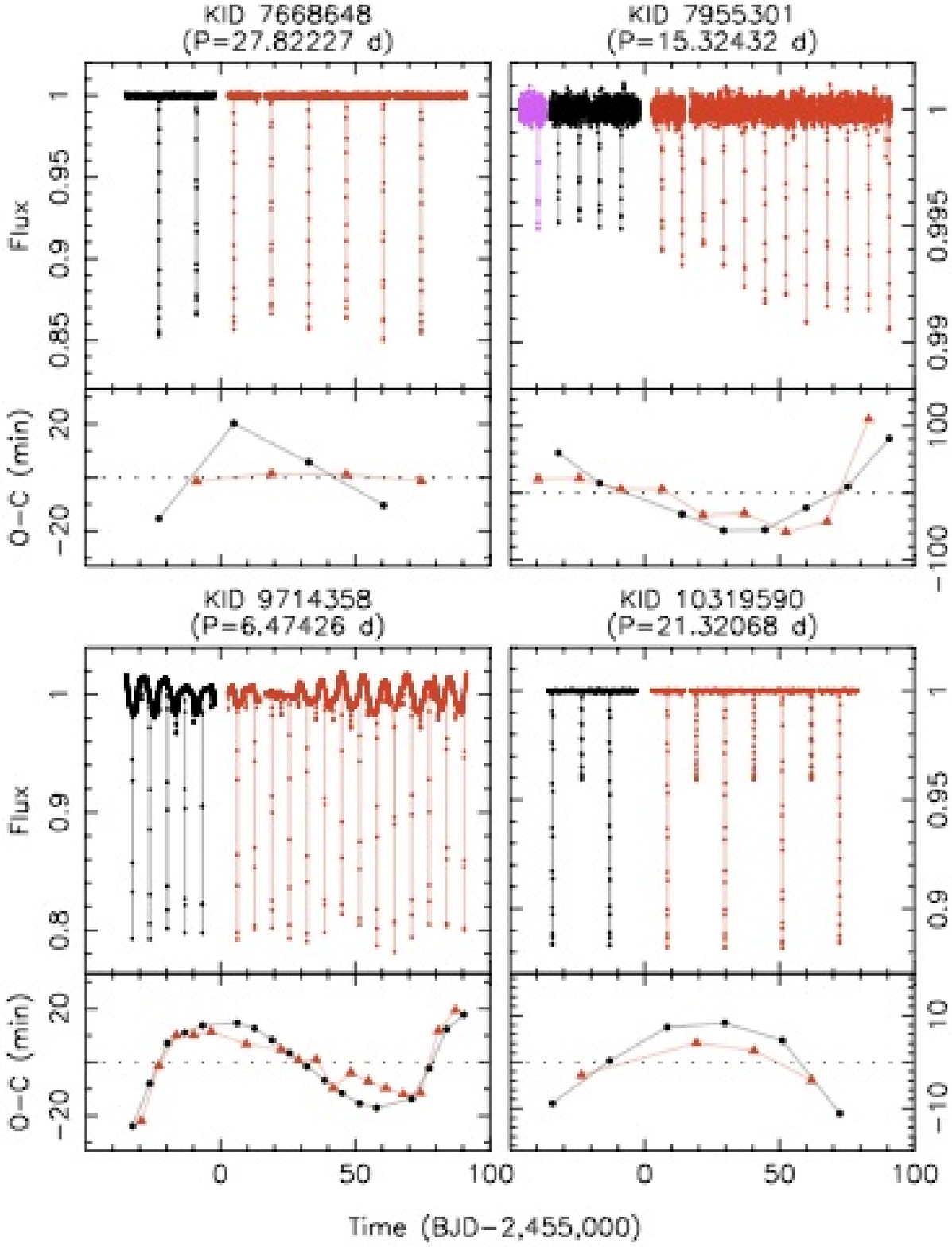}
\caption{EBs with significant signals in the O-C diagram.
The normalized light curves are shown in the upper parts of each panel
where the different colors correspond to data from different quarters,
and the O-C diagram with curves for the primary (filled circles) and
secondary (filled triangles) eclipses are shown in the lower parts of
each panel.
\label{newupdatefig01b}}
\end{figure}

At the time of this writing, about half of the sample has been completed.  The
O-C diagrams were inspected visually, and eight cases where the O-C diagram
has a significant signal through Q2 were identified, see Figs.~\ref{newupdatefig01a}
and \ref{newupdatefig01b}.  KID\,5771589 and KID\,7955301 have changes in
their O-C diagrams of more than 100 minutes.  Most of the others have changes
of 20 to 40 minutes.  It seems unlikely that such large changes in the O-C
diagram over such short times ($\approx 125$ days) can be caused only
by light travel time effects.
We also note that the timescale for apsidal motion is much longer than the
variations seen here.
Hence, each of these EBs is most likely interacting with
a third body.

The list of eight systems includes two of the four EBs with
tertiary eclipses, namely KID\,6543674 and KID\,7289157.  KID\,6543674 has a
modest sized, but coherent signal in both the primary and secondary curves.
On the other hand, the secondary eclipses of KID\,7289157 show roughly a 10
minute O-C variation, whereas the primary eclipses are consistent with a
constant period.  In a similar fashion, both KID\,10319590 and KID\,7668648 have
O-C variations of the primary eclipses that are a bit different than the O-C
variations of the secondary eclipses, although the number of events in each is
not that large.  If confirmed with more data, these period differences between
the primary eclipses and secondary eclipses would almost certainly be a sign
of a dynamical interaction with another body.

Finally, KID\,7955301 seems to have a systematic change in
the eclipse depth. One should be cautious when
interpreting changes in the eclipse depth from quarter to quarter (\S\ref{subsec:qtqsystematics}).
If the eclipse depths really are getting deeper in KID\,7944301,
then that fact should become increasingly evident as more data
become available.

\section{Updated Parameters}

\subsection{Potential Quarter-to-Quarter Systematics\label{subsec:qtqsystematics}}

\begin{figure}
\epsscale{0.8}
\plotone{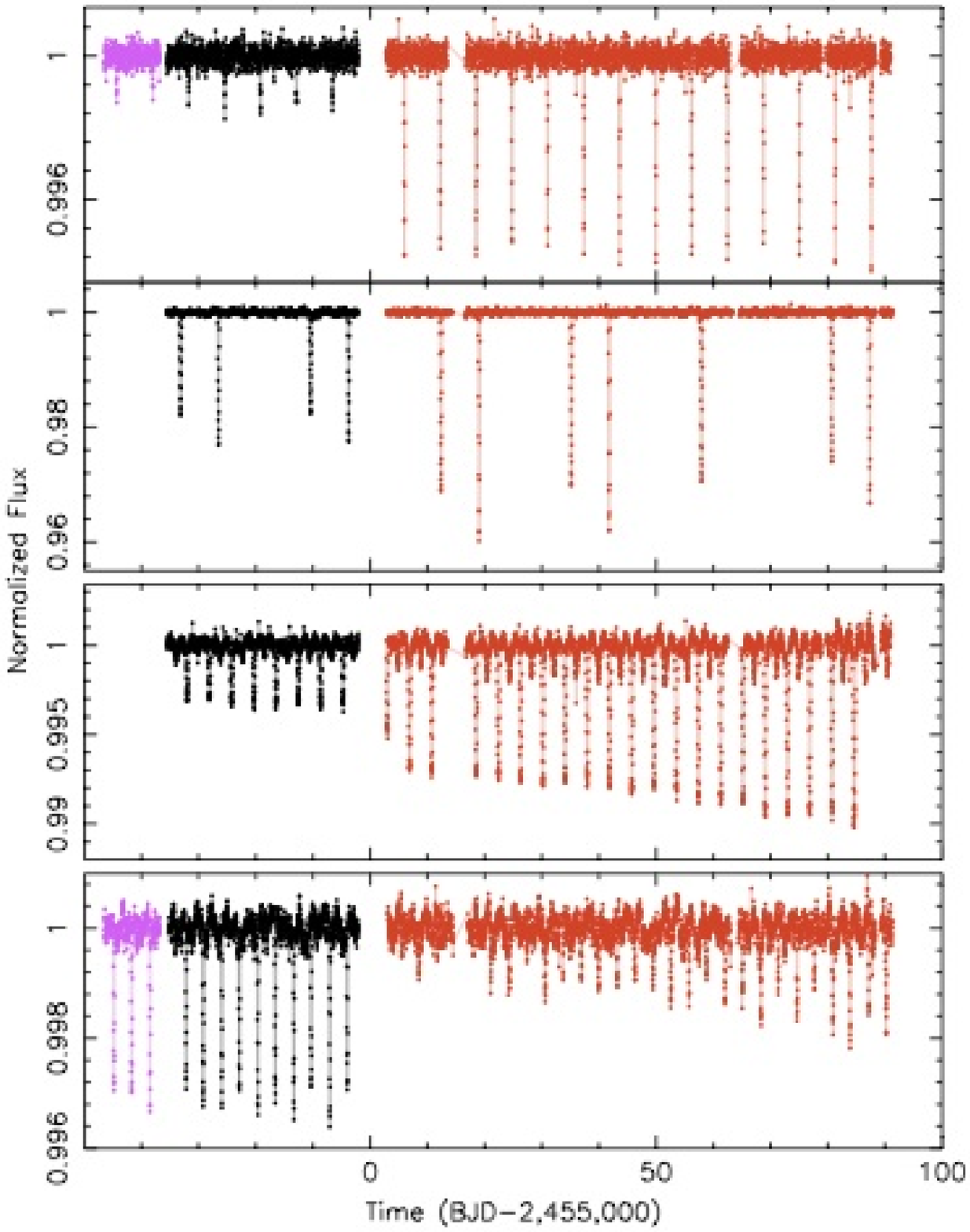}
\caption{Examples of EBs suffering from systematic errors owing to 
changes in the contamination level from quarter to quarter.
To highlight the point, the data from different quarters are shown by
different colors.
The EBs are, from top to bottom, KID\,8255058 ($P=6\fd28383$),
KID\,10491544 ($P=22\fd7729$), KID\,4474645 ($P=3\fd98084$),
and KID\,4073730 ($P=6\fd27129$).
\label{updatefig02}}
\end{figure}

In order to keep its solar panels aligned with the Sun, the \kepler\
spacecraft must roll by 90\degr\ four times a year.  The spacecraft
orientation at a given angle defines ``quarters'', and the spacecraft had a
different orientation during Q2 than it had during Q0/1 (there was no roll between Q0 and Q1).
As a result the
stars are on different CCDs during Q2 compared to Q0/1.
In addition, the
optimal photometric apertures may be quite different from quarter
to quarter particularly for faint stars, and if an EB is in a crowded field,
the amount of blending with other sources
may likewise change from quarter to quarter.
We have found several cases with
abrupt changes in the eclipse depth between Q1 and Q2, and these are
almost certainly caused by differences in the contamination levels.
Figure~\ref{updatefig02} shows four such cases.
These are all clearly EBs, but the
eclipse depths of a few percent or less indicates a high level of
contamination.
Three of these have been shown to be blends
where the targeted object was not the EB (see \S\ref{sec:blended} above) and have been retargeted beginning
with Q8.
Users of {\em Kepler} data are urged to use caution when
combining data crossing quarter boundaries.

\subsection{Data Detrending\label{sec:detrend}}

One of the main issues that prevents reliable catalog-wide EB light curve
fitting is variability in the baseline flux level.
This may be the result of either systematic effects
(focus drifts, safe modes, etc.~-- see \citealt{dr7}), intrinsic stellar
variability (chromospheric activity, pulsations), or extrinsic contamination
by third light (a variable source that contributes light in the aperture of
the object of interest).  The main Kepler pipeline delivers two types of
photometric data: \emph{calibrated} and \emph{corrected}.  Calibrated data are
obtained by performing pixel-level calibration that corrects for the bias,
dark current, gain, non-linearity, smear and flat-field, and applies aperture
photometry to reduced data.  Corrected data are the result of Pre-search Data
Conditioning (PDC) that corrects degraded cadences due to data anomalies and
removes variability to make the targets suitable for planet transit detection
\citep{jenkins:2010b}.  Since the PDC detrending is optimized for planet
transits, its effects on eclipsing binary data are found to be adverse in a
significant fraction of all cases (cf.~the discussion in Paper I, \S2).  That
is why we use only calibrated data\footnote{Specifically, in the FITS tables the calibrated data are in the {\tt `ap\_raw\_flux'}
and the PDC data are in the {\tt `ap\_corr\_flux'} columns.}
in our analyses. As a consequence, the
detrending of data needs to be done as part of our processing pipeline.

The basis of the implemented data detrending algorithm is a least squares
Legendre polynomial fit of order $k$ to the data.  The initial fit takes all
data points into account. Since we want to fit the baseline, we sigma-clip
data points to the asymmetric interval $(-1\sigma, 3\sigma)$: any points that
are $1\sigma$ below the fit and $3\sigma$ above the fit are discarded.  We
re-fit the polynomial to the remaining data points and perform the next
clipping iteration.  The process continues until no data points are clipped.
The fitted polynomial approximates the variable baseline and we divide the
observed data by the polynomial value at each cadence.
This results in a normalized, detrended
light curve that is subsequently phased with the respective ephemeris and
passed to the modeling engine {\tt ebai}.

\begin{figure}
\epsscale{0.61}
\plotone{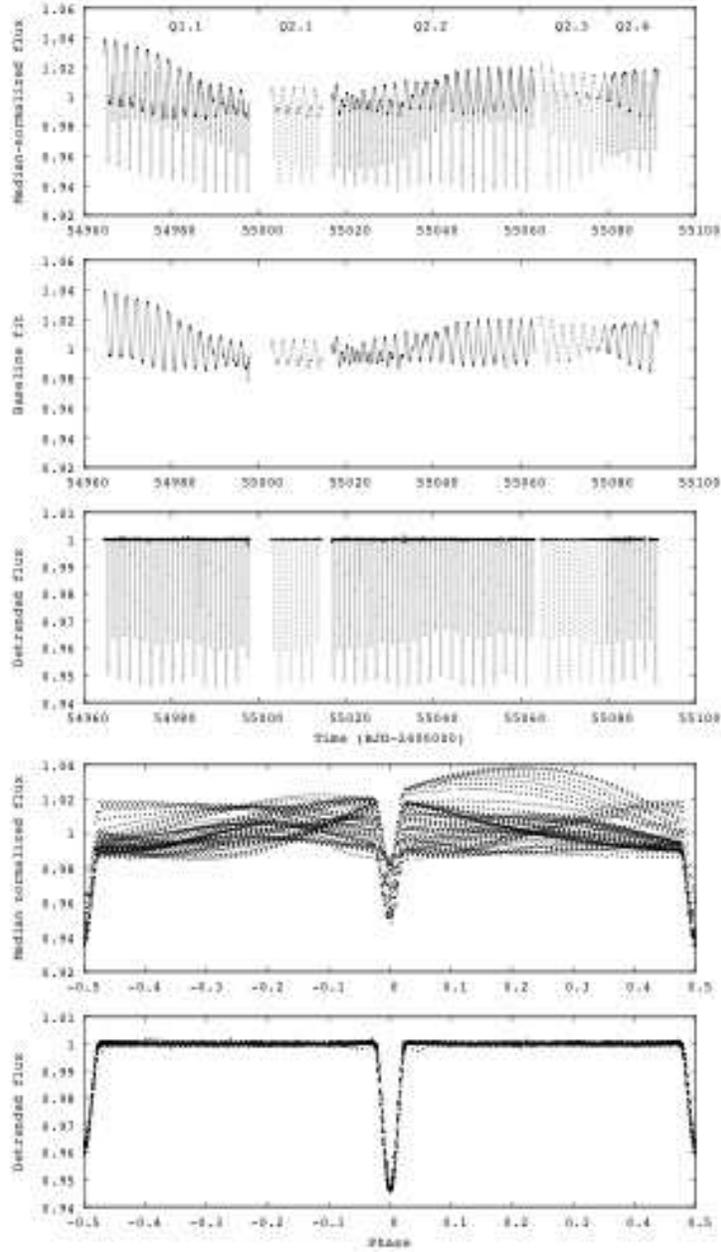}
\caption{
\label{fig:detrending}
An example of the detrending algorithm performance on a detached binary
KID\,12506351 with a strongly variable baseline.  The set is split into 5
parts at the predictable discontinuities: Q1 and 4 sub-parts of Q2: Q2.1,
Q2.2, Q2.3 and Q2.4.  The successive panels show (top-to-bottom): the
median-normalized light curve, the fitted baseline solution, the detrended
light curve, the original phase curve, and the detrended phase curve.}
\end{figure}

Multi-quarter data present a challenge for this approach because of the
discontinuities in the light curves caused by anomalies and random cosmic ray
events.  To account for predictable discontinuities (documented in
\citealt{dr7}), we split the whole data sequence into parts with boundaries at
each discontinuity and detrend each part separately.  The detrending
polynomial order $k_i$ for part $i$ is computed automatically from the number
of data points so that $k_i = (N_i/N) k$, where $N$ is the total number of
data points and $k$ is the suitable polynomial order for the whole data span.
Our attempts to determine the detrending order $k$ automatically had limited
success, so for the most part a manual inspection was
used. Fig.~\ref{fig:detrending} depicts a detrending example of order $k=150$
for KID\,12506351 -- a case of a detached binary with a strongly variable
baseline due to significant chromospheric activity.

To facilitate modeling of phased light curves (or even make it
possible in some cases), aggressive detrending needs to be applied to
the data. This is most notable for chromospherically active stars where
spot modulation causes a significant baseline variability with an
amplitude of the same order as the eclipse depths. In such cases the
detrending is likely to adversely remove variability due to ellipsoidal
variations as well. In order to prevent the negative effects of
detrending on eclipses, we limit the largest fitting polynomial period
to the orbital period and warn that, despite our best effort, processing
artifacts are likely injected into the most variable data-sets and
detailed manual detrending remains necessary.

It is very difficult to automatically detrend random discontinuities in EB
data mostly because eclipses in detached binaries \emph{are} discrete
discontinuities.  At this time we ignore such discontinuities since their
impact on the light curve solution is seldom critical.  However, we are in the
process of implementing a feedback loop from the part of the pipeline that
phases the data: the eclipses are detected and removed, and the data whitened
in this way are subject to a discontinuity search. If a discontinuity is
found, a new breakpoint is added and the set is divided into new subsets.  The
first results seem promising, but further testing and validation is required
before this detrending approach is implemented into the pipeline.

\section{Catalog Analysis\label{sec:analysis}}

\begin{figure}
\plotone{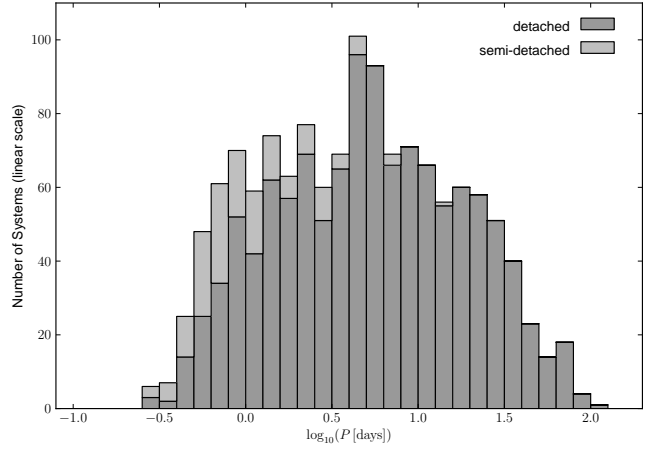}
\caption{Period distribution of the detached (darker grey) and semi-detached (lighter grey)
eclipsing binary systems in the \Kepler\ Q1+Q2 data set.
The baseline is 125 days ($\log P=2.1$).\label{fig:detached}}
\end{figure}

\begin{figure}
\plotone{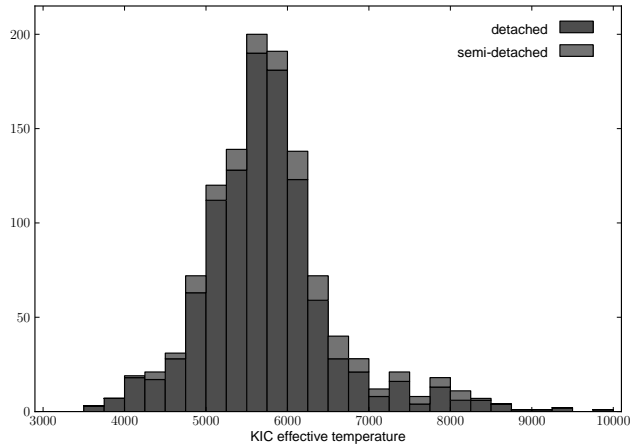}
\caption{Distribution of detached EBs with \Kepler\ Input Catalog effective temperature.
\label{fig:tempDetached}}
\end{figure}

\begin{figure}
\plotone{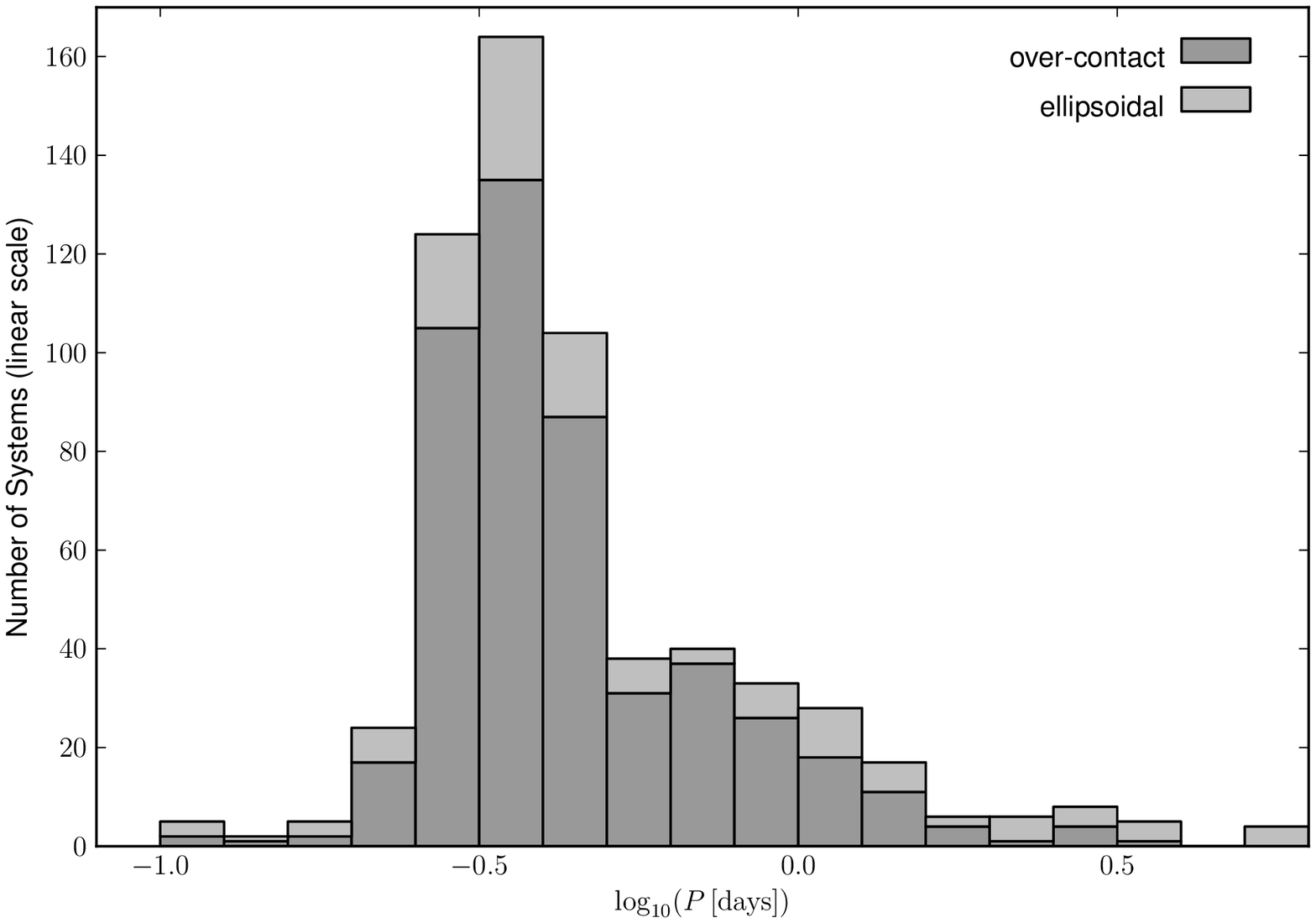}
\caption{Period distribution of the over-contact and ellipsoidal systems in the second \Kepler\  data release.
\label{fig:overcontacts}}
\end{figure}

\begin{figure}
\plotone{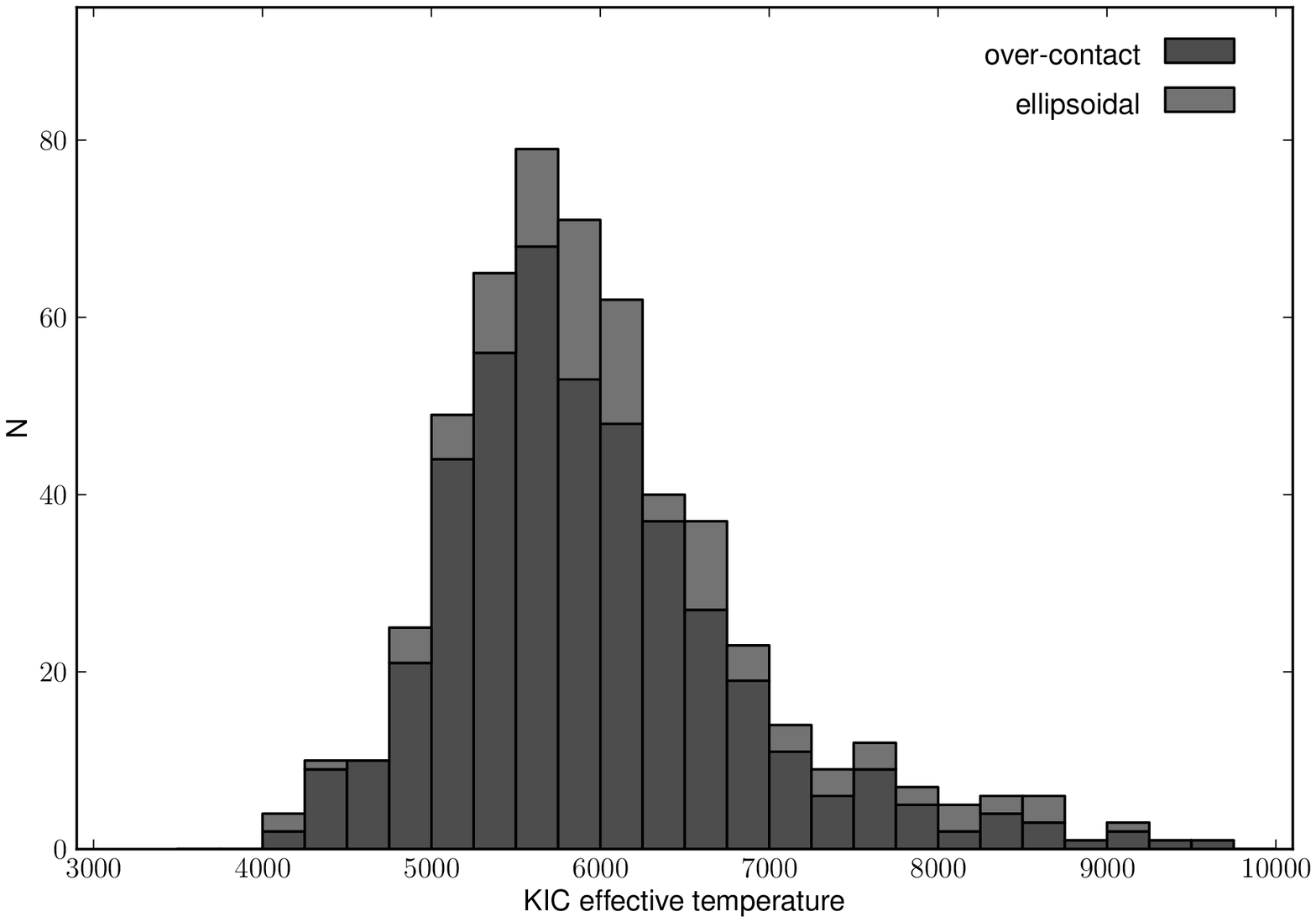}
\caption{Distribution of over-contact and ellipsoidal systems with KIC effective temperature.
\label{fig:tempContact}}
\end{figure}

In this section we provide an updated look at several distributions of the
eclipsing binaries within the \Kepler\ field-of-view.
In Paper I we showed the distribution of EBs as a function of their orbital
period stacked by their morphological type.  The short \Kepler\ Q1 baseline of
34 days limited the analysis to periods less than 25 days.
Here we have re-plotted the distributions in $\log$-Period and examine the distributions with temperature
after dividing the EBs into two subsets, the detached
and semi-detached systems in Figs.~\ref{fig:detached} and \ref{fig:tempDetached}, and the over-contact and ellipsoidal systems in
Figs.~\ref{fig:overcontacts} and \ref{fig:tempContact}.

There is a distinct fall-off in the number of detached systems at both shorter, $P\la0\fd8,$ and
longer, $P\ga45^{d}$, periods.
The short fall-off is due to the relatively small number of low-mass systems
in the magnitude limited \Kepler\/ target catalog.
Such EBs have as components late-K and M dwarf stars
that would be expected to populate this part of the distribution.
In Fig.~\ref{fig:tempDetached}, we show the distribution of detached EBs from the catalog with their
effective temperature as listed in the \Kepler\ Input Catalog.
This shows the sharp decline in the number of systems with decreasing temperature,
again indicative of the same selection effect.
We caution that the effective temperatures in the KIC, which apply to single stars,
have large uncertainties ($\pm 300$ K or more).  Also, no correction has been applied
for binarity which may lead to an underestimate of the temperature of the hotter (usually primary)
component.
The fall-off at periods longer than $\sim45$ days indicates where incompleteness begins to become significant.
Another feature in Fig.~\ref{fig:detached} is the possible excess of detached systems with periods
near 5 days although it is not clear how significant it is.

The log-period distribution of overcontact systems (Fig.~\ref{fig:overcontacts})
has a prominent peak near $-0.5$ ($\sim$0.3 days)
and a smaller, perhaps broader component centered about $-0.2$ ($\sim$0.65 days).
This is suggestive of two separate populations.
In Fig.~\ref{fig:tempContact} we show the KIC temperature distribution for the contact systems.
Interestingly, this distribution has a slight skew towards higher temperatures whereas the
temperature distribution of the detached systems has a slight skew to lower temperatures.

\begin{figure}
\plotone{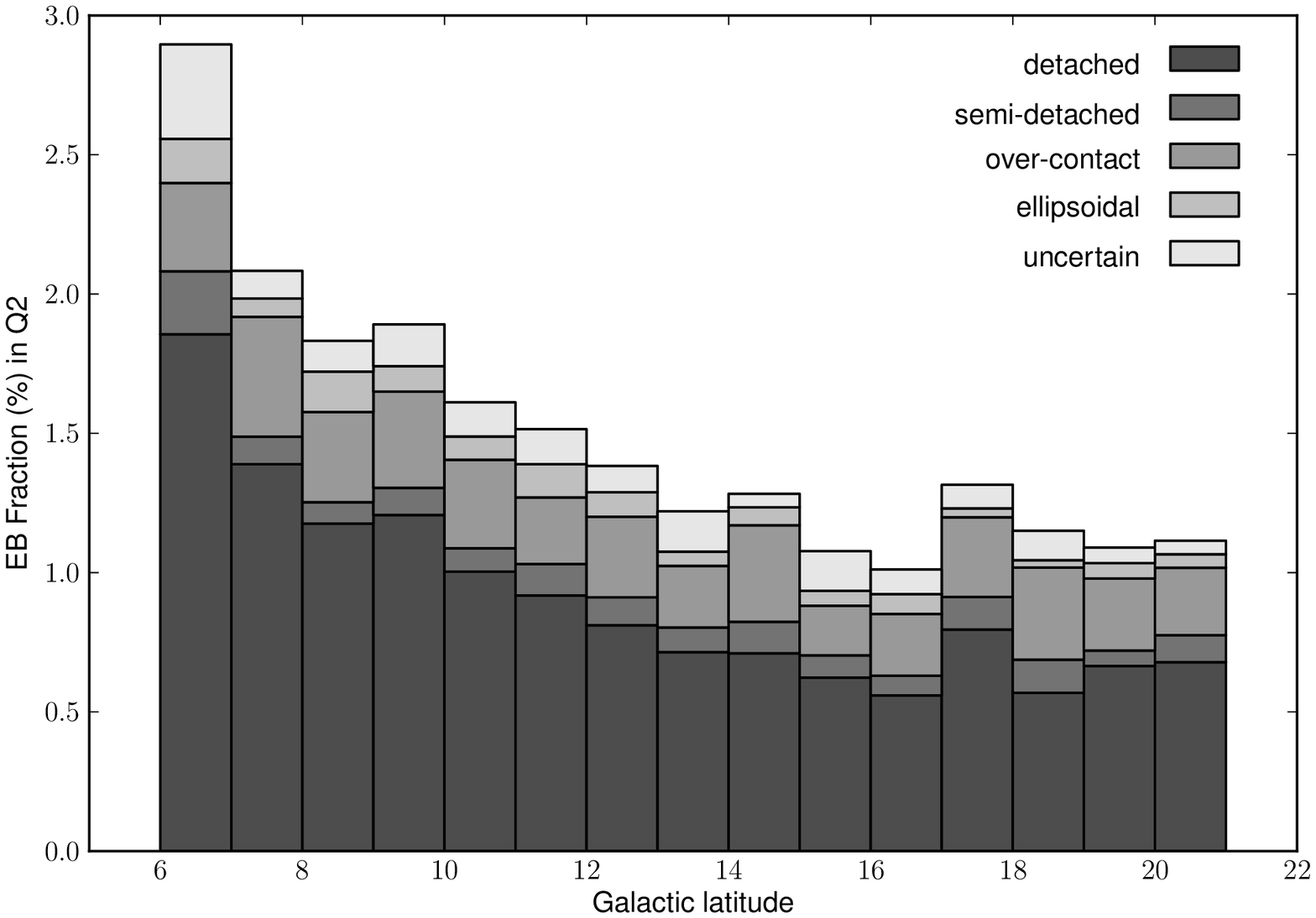}
\caption{The percentage fraction of Q2 targets classified as EBs in each
  $1^\circ$ strip of galactic latitude within the \Kepler\ FOV.  Different
  morphological types are indicated in grey-scale.  The first and last bin are
  affected by the small number of objects in the strips containing the FOV
  edges.
\label{fig:latdist}}
\end{figure}

Figure \ref{fig:latdist} is an update of Fig.~12 in Paper I.
It shows a variation in the fraction of targets that are eclipsing binaries with galactic latitude.
In general, there is an increase in the EB fraction due to the
increase in the number of systems in the updated catalog.

There is a nearly uniform distribution in the eclipsing binary fraction with latitude for the short-period,
interacting overcontact and semi-detached systems, and the ellipsoidal variables.
This indicates a large scale-height perpendicular to the galactic plane for these EBs suggestive of an older population.
In contrast, the detached systems become an increasingly larger fraction of the targets at lower galactic latitudes.
While some of this may be due to increased crowding resulting in a larger number of blends,
it also points to a greater concentration of detached systems towards the galactic plane,  implying 
that these EBs are, on average, younger.

Finally, at the higher galactic latitudes, the total eclipsing binary fraction is observed to flatten out at
$\sim$1.1--1.2\%, essentially the same fraction found in Paper I.

\section{Summary}

The revised catalog of eclipsing binaries contains 2165 entries: 1261
detached, 152 semi-detached, 469 overcontact, 137 ellipsoidal variables and 147
uncertain or unclassified systems. All new entries have been subjected to the same level of
scrutiny as the initial catalog targets: the periods were determined by using
\verb|ephem| and \verb|sahara| tools (Paper I) and the folded light
curves examined for any clear non-EB signatures. All ambiguous cases were
flagged as uncertain (\verb|UNC|) and require further validation.

This revision of the catalog contains a new column \verb|Source|.
It tracks the origin of the added target:
\verb|CAT| if it appeared in the first catalog release, \verb|Q1HB| if it was
held back at the time of the initial release but is now public, \verb|KOI| if
it is a rejected Kepler Object of Interest due to the detected EB signature,
and \verb|NEW| if it was a newly discovered EB.

An online version of the catalog is maintained at
\anchor{http://keplerEBs.villanova.edu}{http://keplerEBs.villanova.edu}.
This catalog lists \kepler\ ID,
morphology, ephemeris, principle parameters, and figures with both time domain
and phased light curves of each system.
It is recommended that anyone wishing to use \kepler\ data for any of these systems consult the 
updated {\it Data Release Notes}
for quarters Q0 and Q1 (first data release), and
Q2 (second data release) that are available at the MAST
website\footnote{\url{http://archive.stsci.edu/kepler/release\_notes/release\_notes5/Data\_Release\_05\_2010060414.pdf}}$^,$\footnote{\url{http://archive.stsci.edu/kepler/release\_notes/release\_notes7/DataRelease\_07\_2010091618.pdf}}.

\acknowledgments

This work is funded in part by
the NASA/SETI subcontract 08-SC-1041 and NSF RUI AST-05-07542.
Doyle and Slawson are supported by the {\it Kepler} Mission Participating
Scientist Program, NASA grant NNX08AR15G.  Welsh and Orosz
acknowledge support from the Kepler Participating Scientists Program via NASA
grant NNX08AR14G.  We thank the following graduate and undergraduate students
at San Diego State University who assisted in measuring the ephemerides:
Gideon Bass, Mallory M. Vale, Michael B. Brady, and Camilla Irine Mura
(visiting from the Universita degli Studi di Pavia, Italy).  We gratefully
acknowledge the use of computer resources made available through Research
Experiences for Undergraduates (REU) grant AST-0850564 to San Diego State
University from the National Science Foundation.

All of the data presented in this paper were obtained from the Multimission Archive at the Space Telescope Science Institute (MAST).
STScI is operated by the Association of Universities for Research in Astronomy, Inc., under NASA contract NAS5-26555.
Support for MAST for non-HST data is provided by the NASA Office of Space Science via grant NNX09AF08G and by
other grants and contracts.
 
Funding for this Discovery Mission is provided by NASA's Science Mission Directorate.
 
{\it Facilities:} \facility{Kepler}

\end{document}